# Controlled dynamic screening of excitonic complexes in 2D semiconductors

Andrey R. Klots[1], Benjamin Weintrub[1,2], Dhiraj Prasai[3], Daniel Kidd[1], Kalman Varga[1], Kirill A. Velizhanin[4], Kirill I. Bolotin[1,2]

[1]Department of Physics and Astronomy, Vanderbilt University, Nashville, TN-37235, USA
[2]Department of Physics, Freie University, Berlin-14195, Germany
[3]Interdisciplinary Graduate Program in Materials Science, Vanderbilt University, Nashville, TN-37234, USA
[4]Theoretical Division, Los Alamos National Laboratory, Los Alamos, NM-87545, USA

**Abstract.** We report a combined theoretical/experimental study of dynamic screening of excitons in media with frequency-dependent dielectric functions. We develop an analytical model showing that interparticle interactions in an exciton are screened in the range of frequencies from zero to the characteristic binding energy depending on the symmetries and transition energies of that exciton. The problem of the dynamic screening is then reduced to simply solving the Schrodinger equation with an *effectively frequency-independent* potential. Quantitative predictions of the model are experimentally verified using a test system: neutral, charged and defect-bound excitons in two-dimensional monolayer $WS_2$, screened by metallic, liquid, and semiconducting environments. The screening-induced shifts of the excitonic peaks in photoluminescence spectra are in good agreement with our model.

**Introduction.** Excitonic complexes (EC) including excitons, trions, and biexcitons are many-body bound states of electrons and holes that can be viewed as solid state analogs of atoms and molecules. Many fundamental atomic physics phenomena such as Bose-Einstein condensation, the Lamb shift, and the fine structure are also observed in ECs[1-3]. One of the key differences between ECs and atomic systems is the size – nanometers for ECs and Angstroms for atoms. While electric fields inside atoms are not perturbed by the environment, the fields in much larger ECs propagate into the surrounding medium and are screened by it. The dielectric properties of the environment can often be adequately described by a *dielectric constant*, $\varepsilon$. In that case, the EC binding energy, $E_{bind}$, can be determined by solving the Schrodinger equation with screened interaction potential, $V$, calculated from the Poisson equation. Many realistic dielectrics, however, are characterized by a *dielectric function*, $\varepsilon(\omega)$, with pronounced frequency-dependence. In that much more complex but experimentally relevant case[4-6], screening becomes *dynamic*, i.e. frequency-dependent. The following question arises naturally: how does one calculate the EC binding energies for frequency-dependent environments?

Effects of dynamic screening are especially interesting in two-dimensional semiconductors from the group of transition metal dichalcogenides (TMDCs). These materials feature a gamut of tightly-bound ECs with binding energies as large as 0.7eV[7,8]. The screening of the ECs, either by their microenvironment[5,9] or by free carriers[10], is especially strong due to the atomic thickness of TMDCs. So far, screening in TMDCs has been modeled as static with the dielectric constant taken either at zero[4,5] or optical[4,11,12] frequencies. While this approach is justified for some systems, for others it may lead to large errors. Although there have been no attempts – to the best of our knowledge – to examine dynamic screening of ECs in TMDCs, theoretical approaches have been developed for conventional semiconductors[13-16]. Unfortunately, these approaches rely on precise knowledge of properties of specific materials and/or require numerical solution of the Bethe-Salpeter equation, and hence are impractical for many realistic systems.

In this work, we develop an analytical model providing intuitive understanding of the screening process. We show that even in the case of dynamic screening, EC binding energies can still be calculated using dielectric functions and screened interaction potentials taken at a certain effective frequency that depends on EC symmetries. We experimentally test the model by studying ECs in monolayer TMDCs coupled to metallic, semiconducting, and liquid environments with frequency-dependent dielectric functions.

**Setting up the problem.** The EC is a system of electrons ($e$) and holes ($h$) bound by an electric field, e.g. neutral exciton ($e+h$), charged exciton ($2e+h$ or $e+2h$, also known as trion), defect-bound exciton (modeled as a trion with one particle being static), etc. We start with a simple semiclassical model of an exciton: two oppositely charged particles revolving around each other inside a homogeneous electrically polarizable medium. In the symmetric case of equally massive particles, $m_e=m_h$, an electron and a hole revolve around their common center of mass with a frequency $\omega_{rot}$. The combined electric field of the particles and hence the polarization of the medium oscillate at the same frequency $\omega_{rot}$. In the opposite asymmetric case, $m_h>>m_e$, the hole is static while the electron revolves around it. Correspondingly, the total electric field created by the charges will have both static and time-dependent components (see Supplementary Information S1). Thus, frequencies relevant for screening of interparticle interactions should depend on EC symmetries in addition to the characteristic frequency $\omega_{rot}$ and related binding energy $E_{bind} \sim \hbar\omega_{rot}$.

We now approach the problem of dynamic screening analytically. Let EC eigenvectors, $|S\rangle$, and eigenenergies, $E_S$, be the solutions of the Schrodinger equation with a frequency-independent interparticle interaction potential. The screening becomes dynamic due to medium excitations, $j_{med}$, such as plasmons or phonons. The corresponding correction to the EC ground state energy can be obtained using the second-order perturbation theory:

$$\Delta E_0 = -\sum_{S,j} \frac{\left|\langle S|\langle j_{med}|H_{int}|0_{med}\rangle|0\rangle\right|^2}{E_{S0}+E_{j0}}. \tag{1}$$

Here, the perturbation $H_{int}$ describes Coulombic interactions of the EC with the medium and the summation is over all possible states of the EC and of the environment. Later we show that while exact expressions for $|j_{med}\rangle$ and $H_{int}$ depend on the structure of a particular solid state system and can be quite complex, knowing their explicit form is not necessary for calculating (1). The multi-index $S = \{n, q\}$ consists of an index $n$ describing internal excitations of the EC (Rydberg series) and the total momentum $q$ of the EC as a whole. Finally, $E_{S0}$ and $E_{j0}$ are the transition energies between ground and excited states of the EC and the medium respectively. Evidently, $\Delta E_0$ depends on EC transition energies $E_{S0}$ starting with $E_{00} = 0$.

Instead of burdensome expressions for $|j_{med}\rangle$ and $H_{int}$, an experimentally accessible dielectric function can be used to describe the dielectric response of the medium. Then, the Poisson equation with medium dielectric constants evaluated at each frequency $\omega$ yields the dynamically screened $\omega$-dependent interaction potential, $V(\omega)$. We note that $V(\omega)$ may have a complex spatial or, equivalently, momentum($q$)-dependence. For example, in a two-dimensional material sandwiched between two dielectrics interparticle interactions are described by the Keldysh potential[17]. We, however, do not write this $q$-dependence explicitly, since our main focus is the frequency-dependence of interactions. The interaction potential $V(\omega)$ consists of an unperturbed frequency-independent potential[1], $V_0$, and a complex-valued dynamic term, $V_s(\omega) = V_s'(\omega) + iV_s''(\omega)$, henceforth referred to as the *screening potential*. Treating $V_s(\omega)$ as a perturbation potential, we can rewrite equation (1) without explicit involvement of $j_{med}$ [13,14]:

$$\Delta E_0 = -\frac{1}{2}\frac{1}{A}\sum_S |\rho_{S0}|^2 \tilde{V}_s(E_{S0}/\hbar). \qquad (2)$$

Here $A$ is the crystal volume, $\tilde{V}_s(E_{S0}) = 2\pi^{-1}\int_0^\infty V_s''(\omega)(\omega + E_{S0}/\hbar)^{-1} d\omega$ [13], and $\rho_{S0} = \langle S|\rho(q)|0\rangle$ is a charge density operator in momentum space "sandwiched" between EC ground and excited state-vectors (See Supplementary Information S1). By analogy with transition dipole moment, $\rho_{S0}$ can be also called the transition charge density. Throughout the paper we use unitless elementary charge $e = 1$.

**Relevant screening frequencies.** While it is possible to numerically compute $\Delta E_0$ from equation (2), such calculations require evaluation of wavefunctions for all of the EC excited states. This is complex even for neutral excitons and impractical for larger ECs. However, we can further simplify equation (2) by using the general properties of $\tilde{V}_s$ and $\rho_{S0}$ (see Supplementary Information S1):

---

[1] calculated at a frequency where the dielectric function is approximately constant.

(a) <u>The frequency-integral</u> $\tilde{V}_s$ can be expressed, using the Kramers-Kronig relations, as frequency-smoothened real part of the screening potential, $V'_s$:

$$\tilde{V}_s(E_{S0}) = \int_{-\infty}^{\infty} f(\ln E_{S0}/\hbar - \ln\omega) V'_s(\omega) d\ln\omega, \tag{3}$$

where $f(x) = 2\pi^{-2} x/\sinh x$ is a normalized bell-shaped distribution function with a vanishing mean value and standard deviation of ~2. According to (3), $\tilde{V}_s(E_{S0})$ can simply be approximated by a real part of the screening potential $\tilde{V}_s(E_{S0}) \cong V'_s(E_{S0}/\hbar)$, provided that $V'_s(\omega)$ is a slow-varying function of frequency. This approximation is valid for many real media[18-21] and is used henceforth to simplify derivations.

(b) <u>The transition charge density</u> created by an electron and a hole – as can be shown analytically – vanishes if $|0\rangle$ and $|S\rangle$ are both symmetric with respect to exchange between electron and hole coordinates $r_e \leftrightarrow r_h$. In the case of such *symmetric transition*, the contributions to $\rho_{S0}$ from an electron and a hole are equal in magnitude and opposite in sign and therefore cancel each other out. Thus, only the *asymmetric* transitions contribute to the sum in (2). This condition is analogous to selection rules in atomic physics. As a result, the minimal value, $E_{\min}$, of the transition energy $E_{S0}$, contributing to the sum in (2), is the *energy difference* between *the ground state* and *the lowest asymmetric state*. The summation in equations (1,2) also has a characteristic upper-bound cutoff energy of the order of the EC binding energy, $E_{\max} \sim |E_{bind}|$ [22,23]: due to decreasing overlap between $|0\rangle$ and $|S\rangle$, the terms corresponding to transition energies above that cutoff quickly decay with increasing $E_{S0}$, allowing the sum in (2) to converge. Thus, only some of the lower-energy terms in (2) effectively contribute to $\Delta E_0$.

(c) <u>The summation in equation (2)</u> can be further simplified by replacing the frequency-dependent function $V'_s(E_{S0}/\hbar)$ by a frequency-independent mean value $V'_s(E_{eff}/\hbar)$ where the effective energy, $E_{eff}$, is a constant lying between the lower and upper energy bounds, $E_{\min} < E_{eff} < E_{\max}$. This assumption of static screening allows one to treat the EC as a set of particles interacting via frequency-independent potential $V_0 + V'_s(E_{eff}/\hbar) = \text{Re}V(E_{eff}/\hbar)$. In this case, the perturbed ground state energy is

$$E_0 + \Delta E_0 = \langle 0|T + \frac{1}{2}\sum_{j,k} Q_j Q_k \left(V_0(r_{jk}) + V'_s(r_{jk}, E_{eff}/\hbar)\right)|0\rangle, \tag{4}$$

where $Q_j$ is the charge of the *j*-th particle, $r_{jk}$ is the interparticle distance and $T$ is the total kinetic energy of all the particles in the EC.

It is instructive to consider examples clarifying the evaluation of the lower-bound energy $E_{\min}$. In the case of a neutral exciton with equal electron and hole masses[24], the ground state $n=0$ is symmetric[^2]. Then, the energy of the first *asymmetric* transition is $E_{\min} \approx E_{1,0} = E_{n=1} - E_{n=0}$, which typically is of the same order as $|E_{bind}|$ [4]. Other common ECs such as trions, defect-bound excitons or neutral excitons with uneven *e-* and *h*-masses behave differently. Their ground state wavefunctions are inherently asymmetric with respect to $r_e \leftrightarrow r_h$ exchange[7]. The lowest asymmetric transition for such ECs is purely translational (with no change in *n*) with $E_{\min} \to 0$. Realistically, an EC may decay before the medium has enough time to get fully polarized. Hence, the effective $E_{\min}$ is not exactly zero, but is limited by the inverse characteristic lifetime $\sim \tau^{-1}$ of the particles constituting the EC.

Equations (3,4) along with the estimates of $E_{eff}$ constitute our main theoretical result. In (4), we effectively replace the dynamically screening medium by a medium with a static dielectric constant $\varepsilon(E_{eff}/\hbar)$. To enable experimental predictions from (4), we note that the 'diagonal' terms with $k = j$ represent *self-interaction* of each carrier with its image charges. 'Off-diagonal' terms with $k \neq j$ account for screening of *interparticle interactions* (i.e. EC binding). Within simple, but widely used effective-medium approximations for interaction potentials, the calculation of self-energies is very susceptible to small uncertainties in microscopic structure of the investigated system and can even yield divergent results[23]. However, the binding energy, calculated using off-diagonal ($k \neq j$) terms in (4), can still serve as a proxy for evaluating strength of interparticle interactions, screened by the medium with *effective* dielectric constant $\varepsilon(E_{eff}/\hbar)$.

In summary: the range of binding energies of ECs *dynamically* screened by environment with dielectric function $\varepsilon(\omega)$ can be evaluated, to the second order of the perturbation theory, by simply solving the EC Schrodinger equation with the *effective dielectric constants,* obtained from the true frequency-dependent dielectric function evaluated at two limiting frequencies: $\omega_{\min} = E_{\min}/\hbar$ and $\omega_{\max} = E_{\max}/\hbar \sim |E_{bind}|/\hbar$. Binding energies obtained from these two cases are the upper and the lower bounds for the actual EC binding energy. The lower bound depends on the EC symmetry: $E_{\min} \approx E_{1,0} \sim |E_{bind}|$ for symmetric charge-neutral ECs with equal *e/h* masses and $E_{\min} \sim \hbar/\tau$ (inverse lifetime of particles constituting the EC) for asymmetric ECs with unequal *e/h*-masses or non-zero net charge. In some specific cases the problem can be simplified further. For example, in the case of a long-lived exciton with $m_h \gg m_e$, a heavy hole can be effectively treated as static and its field – as

[^2]: For a realistic system of *nearly* equal *e-* and *h*-masses in TMDC, $\rho_{00}$ is proportional to the mass discrepancy between an electron and a hole (2~20%). Hence, $|\rho_{00}|^2$ entering (2) does not exceed ~4% compared to the case of unequal *e/h*-masses.

constant. Such a field, and hence, exciton binding will be screened by the medium only at zero effective frequency $\omega=0$ yielding a static effective dielectric constant $\varepsilon(\omega=0)$. Below we will demonstrate that for many realistic cases, $\varepsilon$ does not change significantly between frequencies $E_{min}/\hbar$ and $E_{max}/\hbar$, which allows us to make experimentally testable predictions regarding screening of EC binding.

**Setting up the experiment.** In order to test the developed theory, we measure the effect of different dispersive environments on binding energies of different types of ECs in a monolayer TMDC. We choose monolayer $WS_2$ as a test bed since this material has a variety of tightly bound ECs[4,8,10,25-27] that produce narrow and well-resolved peaks in photoluminescence (PL) spectra[4,8,10,25,28]. We focus on three prominent excitonic species (Fig.1a):

(a) <u>neutral exciton</u> ($X^0$). It has nearly identical electron and hole masses[7,24] and is symmetric according to our classification. Therefore, interparticle interactions are expected to be screened at an effective energy in the mid-IR range: between the first excited state transition energy of ~130meV[4] and binding energy of ~320meV[4].

(b) <u>trion</u> ($X^-$). This charged state is classified as asymmetric. In the case of trion, we expect screening in the THz range: between ~0.5meV, which corresponds to ~10ps lifetime[29,30], and the binding energy ~30meV[8].

(c) <u>defect-bound exciton</u>[26,27] ($X^D$), treated here as a neutral exciton bound to a static charged impurity[3]. The binding energy of $X^D$ is ~150meV, which agrees with our numerical model described below. Note that the binding energies of $X^D$ and $X^-$ are defined with respect to the energy of a neutral exciton. The electric field of a static charged impurity, binding the exciton, is screened at zero frequency. This situation is similar to the example of a long-lived strongly asymmetric exciton considered above[4]. Therefore, defect-bound excitons are expected to be screened at zero frequency.

To test the dynamic screening of these ECs, we choose the media with qualitatively different dielectric functions in the range of relevant frequencies (Fig.1b):

(i) <u>metallic medium</u>. Two-dimensional semimetal graphene exemplifies a metallic-type dielectric response $\varepsilon \sim \omega^{-2}$. Specifically, $\varepsilon(\omega)$ for graphene is large (>10) for $\omega$ from 0 to THz and is close to 1 in the IR range.

---
[3] Currently, the origin of impurities is not completely clear. However, agreement of the measured $X^D$ binding energy with our numerical modeling (electron+hole+static charge) suggests that defect-bound excitons can be treated as a neutral excitons bound to deep charged defects.
[4] For a defect-bound exciton it is energetically favorable to have an electron highly localized near an impurity (if impurity charge is positive) and hole delocalized. Such a distribution of the density function makes the defect-bound exciton indeed similar to a highly asymmetric neutral exciton.

(ii) <u>liquid medium</u>. We use ionic liquid DEME-TFSI[5], for which $\varepsilon(\omega)$ is large (>10) at sub-GHz frequencies and is insignificant above 1THz.

(iii) <u>semiconducting medium</u>. For semiconductors, $\varepsilon(\omega)$ is roughly constant in a broad range of frequencies. In our experiments, monolayer MoS$_2$ transferred onto our device serves as a semiconducting screening layer with $\varepsilon(\omega)$~15 in IR-to-visible range and ~5 in the sub-THz range.

Figure 1b shows the dielectric functions for each medium along with the frequency ranges (shown as vertical bands) relevant for screening of $X^0$, $X^-$, and $X^D$. The dielectric functions are relatively constant within each band. Summarizing, we expect the binding energy of neutral excitons to be strongly affected by semiconducting but not liquid or metallic environments. For trions, we expect strong screening by metallic environment only. Finally, defect-bound excitons should be affected by metallic and liquid environments. We cannot make a definitive qualitative prediction of the effect of the semiconducting medium on $X^-$ and $X^D$ because, in relevant sub-THz range, MoS$_2$ dielectric constant ($\varepsilon$~5) is neither large (>10) nor small (~1).

**Measurements.** Measurements were performed on monolayer WS$_2$ flakes, exfoliated on Si/SiO$_2$ substrates. Electrostatic gating was used to control the Fermi level and isolate the contribution of free-carrier screening[8,10]. In order to study $X^D$ we induced defects using argon plasma[27]. We begin our measurements by recording PL spectra (532nm, ~20µW laser excitation focused into a ~2µm spot) at $T$=78K for pristine WS$_2$ devices without any material on top (Fig.1c, WS$_2$ device). The well-known peaks in the PL spectra at ~2.06eV (black dashed line), ~2.03eV (blue dashed line), ~1.92eV (green dashed line) are identified as stemming from neutral excitons $X^0$, trions $X^-$ and defect-bound excitons $X^D$ respectively [7,8,10,25,27]. The peak at ~2.02eV observed in some devices (e.g. Fig.1c, pink dashed line) is likely associated with an additional trion state[2,10,31] and is not analyzed further.

We modify the dielectric environment of the WS$_2$ flake by either mechanically transferring[32] monolayer graphene or MoS$_2$ (WS$_2$/metal and WS$_2$/semiconductor device respectively), or dropcasting a layer of ionic liquid (WS$_2$/liquid device). We then re-acquire the PL spectra. We observe large and reproducible shifts of all three excitonic peaks (Fig.1c). Note that environmental factors other than screening (i.e., induced doping, strain, and chemical modifications) may also cause peak shifts[8,10,33,34]. However, as shown below and in the Supplementary Information S3, the observed shifts are too strong to be explained by changes in the doping level. The effects of strain are shown to be weak by comparing PL spectra of transferred heterostructures and naturally grown WS$_2$ bilayers. We also see no evidence of chemical modifications in WS$_2$/liquid devices as observed shifts are reversed by *removing* the ionic liquid. Thus, we interpret observed shifts as originating from the dielectric screening of excitons. To compare these shifts with theory, we extract exciton binding energies for different types of environment. The binding energies of trions and defect-bound excitons are determined as $|E_{bind}(X^{-,D})| = Pos(X^{-,D}) - Pos(X^0)$, where $Pos(X)$ is the energy position of a particular excitonic peak in the PL spectrum. In pristine devices, we observe $|E_{bind}(X^-)| \sim 25\text{meV}$ and $|E_{bind}(X^D)| \sim 140\text{meV}$, close to literature values[8,10,27]. Unfortunately, $|E_{bind}(X^0)|$ cannot be measured directly using absorption or PL spectroscopies as these techniques are unable to directly probe the single-particle electronic bandgap [4,8,35]. We rely on the on the experiments by Chernikov, et al.[4,10] measuring $|E_{bind}(X^0)| \sim 320\text{meV}$ for uncovered Si/SiO$_2$/WS$_2$ devices similar to ours, and showing 1meV red-shift in $Pos(X^0)$ per ~6meV decrease in the exciton binding energy (studied by controlling the binding energy by either varying the number of layers or the carrier density in WS$_2$). These observations allow us to convert the screening-induced shifts of the $X^0$ PL peak position into its effective binding energy.

---

[5] diethyl methyl(2-methoxyethyl)ammonium bis(trifluoromethylsulfonyl)imide

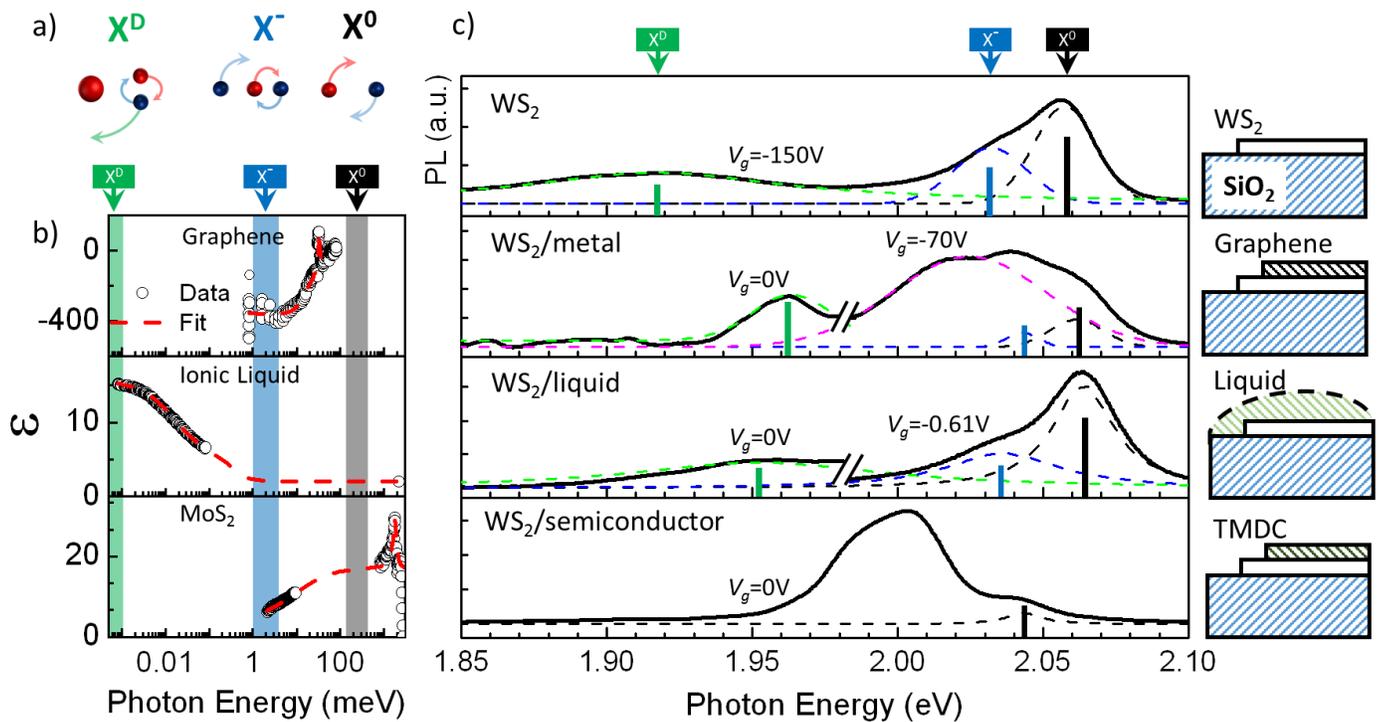

**Figure 1. Effect of environments on WS$_2$ PL spectra. (a)** top: schematic illustrations of $X^D$ (static impurity is in the middle), $X^-$ and $X^0$. **(b)** Dielectric functions of the screening materials: graphene[18], ionic liquid[19,20], and monolayer MoS$_2$[21]. Since experimental dielectric functions are not available for the entire frequency range, we interpolate them using double Lorentzian fitting. **(c)** PL spectra of WS$_2$ in different environments – schematics are on the right. Dashed curves are fitted excitonic peaks. The symbol "//" separates curves obtained from different samples/at different gate voltages. Voltage is shown above each curve. As in-situ gating with ionic liquid is impossible at low temperatures, the data for the WS$_2$/liquid device (right curve) were obtained at 240K and artificially blue-shifted by 40meV to account for thermal shift of the peaks[25].

Figure 2 summarizing the effects of metallic, semiconducting, and liquid environments on the binding energies of $X^0$, $X^-$, and $X^D$ (square symbols) constitutes our main experimental result. The following trends are evident: The extracted binding energy of $X^0$ decreases by 120±40meV (~40%) in the $WS_2$/semiconductor sample. This conforms well with studies performed on bi- and multi-layer TMDCs[4,36,37]. For $X^-$, the binding energy is downshifted by 10±3meV (~30%) due to the presence of graphene. The binding energy of $X^D$ is reduced by 40±20meV (~30%) in presence of both metallic and liquid environments. In all other measured cases EC peak shifts are insignificant within our error bars. These trends agree well with our qualitative predictions. In the case of $WS_2$/metal and $WS_2$/semiconductor samples we could not bring $WS_2$ close to depletion, likely due to strong effects of charge transfer in these heterostructures[38]. Nevertheless, observed shifts exceed possible doping-induced effects: the trion binding energy in presence of graphene becomes as low as 19meV, and the neutral exciton red-shifts to 2.045eV in semiconductor-capped devices. These values are significantly below the energies achieved by doping alone[8,10](see Supplementary Information S3).

**Quantitative comparison with theory.** To further verify our model, we perform quantitative estimates of ECs binding energies (see Supplementary Information S2). We computationally solve the Schrodinger equation for 2- or 3-body systems using a variational approach[39-41] with $e$- and $h$-masses of $0.45m_0$[24,42] and infinite mass for the defect charge. Interparticle interactions are modelled by the Keldysh potential[17] calculated using $WS_2$ and medium dielectric functions taken at effective frequency $\omega$. Upper- and lower-bound estimates for EC binding energies ($E_{bind}(\omega_{min})$ and $E_{bind}(\omega_{max})$) are obtained by setting $\omega$ to $\omega_{min} = E_{min}/\hbar$ or $\omega_{max} = |E_{bind}|/\hbar$ as prescribed by our theoretical model.

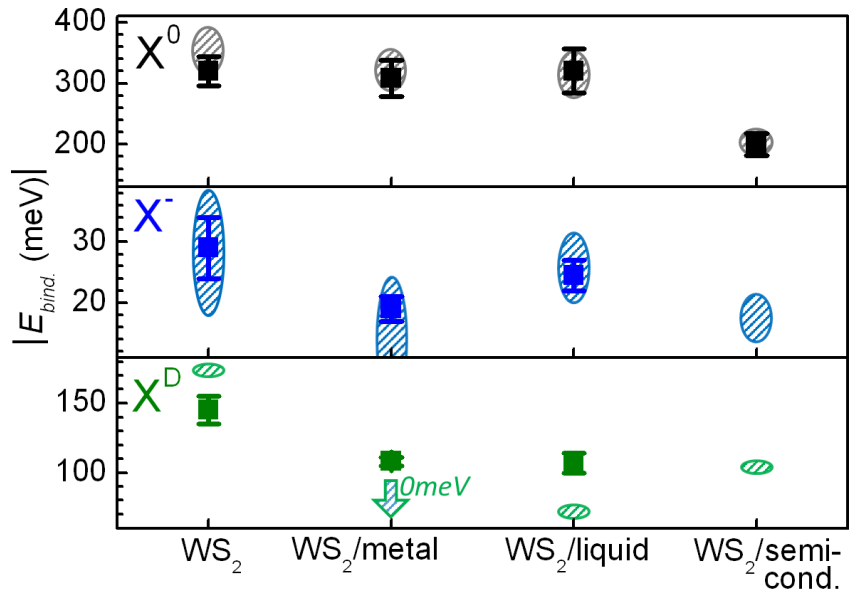

**Figure 2. Summary of experimental and theoretical results.** Square symbols are experimentally observed EC binding energies in presence of different screening materials, while ovals show the range of theoretically predicted values. For both $X^-$ and $X^D$ in $WS_2$/metal devices the calculated energy range starts at zero (shown by downward arrow in the case of $X^D$).

The ranges of theoretical EC binding energies – from $E_{bind}(\omega_{min})$ to $E_{bind}(\omega_{max})$ – are shown as shaded ovals in Fig.2. Observed values of $X^0$ and $X^-$ binding energies are within the theoretically expected range for all media. Shifts of $X^D$, calculated assuming only zero-frequency screening, exceed experimental ones, probably due finite spatial separation between the measured EC and the medium, which is assumed to be negligible in our model. In the case of $X^-$ and $X^D$ in the presence of a semiconductor environment, predicted shifts are too subtle to be experimentally tested with certainty and were not measured as that would require higher accuracy of computational models and measurement techniques. Overall, we believe that this quantitative agreement is remarkable for a minimal model with no free parameters.

**Conclusions.** The theory of excitonic complexes in dynamically-screening media was developed and confirmed experimentally. We obtained the binding energies of dynamically screened ECs by solving the Schrodinger equation with *effectively static* interaction potentials calculated at the *fixed effective* frequency. This frequency depends on the symmetries of the wavefunctions and the binding energies of ECs. The model was tested and confirmed experimentally by using neutral, charged, and defect-bound excitons in monolayer $WS_2$ screened by metallic, semiconducting, and liquid environments. The developed approach is general and can be applied to diverse systems of quasiparticles, interacting via electric fields: including plasmons, excitonic molecules, and polaritons, screened by various media.

Our simple dynamic screening model may help to re-interpret and clarify a wide range of previous experiments where static screening was assumed. For example, the assumption of zero-frequency screening of two-dimensional ECs by liquids ($\varepsilon(\omega=0) \sim 50$) has led to the appearance of outlying data points, overestimation of exciton binding energies[5,43] and underestimation of the effective electron mass by two orders of magnitude[44]. Moderate shifts in exciton energies observed in these experiments are more consistent with screening at optical frequencies, as predicted by our model, where most liquids have $\varepsilon \sim 2$. Another important example is the inconsistency in the reported neutral exciton binding energy in monolayer $MoS_2$, which ranges from 220meV to 660meV depending on the type of measurements and applied models[35,45,46]. The lowest binding energy, 220meV, is obtained by Zhang et *al.*[46] by subtracting the optically measured energy of the excitonic PL peak from the electronic bandgap measured using scanning tunneling spectroscopy. Their measurements were performed using $MoS_2$ samples on a semimetallic graphite substrate. According to our model, excitonic and free-particle states are screened by graphite at different effective frequencies, which yields ~400meV difference in corresponding screening-induced energy shifts. This accounts for the discrepancy between the values obtained by Zhang et *al*. and by others[35,45].

Effects of dynamic screening may also have practical applications. For example, it may be possible to probe frequency-dependent dielectric functions of various microscopic environments by measuring relative shifts of different types of ECs (including EC excited states) that are screened at different effective frequencies. This can be interesting for label-free biodetection or chemical sensing.

**Acknowledgements.** We thank Moshe Harats, Ryan Nicholl, Jason Bonacum and Slava Rotkin for useful discussions and comments. K.I.B. acknowledges support from ONR N000141310299 and NSF DMR 1508433. Samples for this work were prepared at the Vanderbilt Institute of Nanoscale Science and Engineering.

**Author contributions.** AK and KAV developed the theoretical approach. BW, DP and AK fabricated the samples. AK and BW performed the measurements. KB supervised the project. DK, KV performed numerical calculations. AK, KV and KB co-wrote the manuscript with contributions from all authors.

# Controlled dynamic screening of excitonic complexes in 2D semiconductors

Andrey R. Klots[1], Benjamin Weintrub[1,2], Dhiraj Prasai[3], Daniel Kidd[1], Kalman Varga[1], Kirill A. Velizhanin[4], Kirill I. Bolotin[1,2]

[1]Department of Physics and Astronomy, Vanderbilt University, Nashville, TN-37235, USA

[2]Department of Physics, Freie Univerity, Berlin-14195, Germany

[3]Interdisciplinary Graduate Program in Materials Science, Vanderbilt University, Nashville, TN-37234, USA

[4]Theoretical Division, Los Alamos National Laboratory, Los Alamos, NM-87545, USA

## Supplementary Information

# S1. Theoretical approach.

## S1.1. Semiclassical model.

Let us consider dielectric screening of the electric field in a classical model of an EC: two charged particles rotating around each other in a dielectric medium. We assume that the particles are rotating with the frequency $\omega$ and are separated by the distance $\rho$. We first consider the symmetric case of equally massive and oppositely charged (with charge $Q$) particles rotating around the common center of mass. The displacement field $D$ at any point of space $r$ can be expressed as the gradient of the potential created by the particles:

$$D = -\nabla \frac{Q}{\sqrt{|r|^2 + |\rho|^2/4}} \left[ \frac{1}{\sqrt{1 + \frac{|r||\rho|}{|r|^2 + |\rho|^2/4}\cos\omega t}} - \frac{1}{\sqrt{1 - \frac{|r||\rho|}{|r|^2 + |\rho|^2/4}\cos\omega t}} \right]$$

In this case, the displacement field is clearly the odd function of $\cos\omega t$. Hence, $D$ has only odd harmonics with frequencies $(2n+1)\omega$ (here $n \in Z$) and hence, will not have any zero-frequency harmonics.

In the opposite asymmetric case of much heavier positive (for instance) particle, that particle is considered static. The displacement field reads:

$$D = -\nabla \left[ \frac{Q}{\sqrt{|r|^2}} - \frac{Q}{\sqrt{|r|^2 + |\rho|^2 - 2|r||\rho|\cos\omega t}} \right].$$

This expression for displacement field is neither symmetric nor antisymmetric function of $\cos\omega t$. This function has all of the harmonics with frequencies $2n\omega$, including the zero-frequency mode. Therefore, the contribution of the dielectric response of the medium at zero frequency to screening depends on the symmetry of the EC.

## S1.2 Charge density matrix element in the center-of-mass frame.

In this section, we derive the expression for the charge density matrix element and derive equation (2) of the main text. We start with expression for the screening-induced perturbation correction of the ground state energy of the excitonic complex (EC) in the second order of the perturbation theory[1,2]:

$$\Delta E_0 = -\frac{1}{2}\frac{1}{A}\sum_{S,q}|\rho_{0S}(q)|^2 \frac{2}{\pi}\int_0^\infty \frac{\text{Im}V_s(\omega,q)}{\omega + E_{S0}}d\omega. \tag{S1.2.1}$$

Here and further in the text we use atomic units $e = \hbar = 1$. As mentioned in the main text, multi-index $S = \{n,k\}$ consists of $n$ - the index labelling internal excitations of the EC in the center-of-mass frame and $k$ - momentum of the EC as a whole. Fourier-transformed screening potential $V_s(\omega,q)$ depends on wavenumber $q$ and frequency $\omega$. To simplify (S1.2.1), we perform the calculation in the center-of-mass frame. We write the wavefunction of EC with momentum $k$ and internal quantum number $n$ as

$$\psi_S \equiv \psi_{k,n} = A^{-1/2}\exp(ikr_C)\Psi_n(r_1',r_2',...), \tag{S1.2.2}$$

where $r_C$ is the coordinate of EC's center of mass and $\Psi_n$ is the center-of-mass wavefunction. Coordinate $r'_j$ is a center-of-mass (CM) frame coordinate (denoted by the "prime"-symbol) of $j$-th particle. Coordinates in a static/laboratory reference frame will be denoted as $r_1, r_2, ...$ (no "prime"). Charge density of an $N$-particle excitonic complex is written as $\rho(r) = \sum_{j=1}^{N} Q_j \delta(r - r_j)$, where $Q_j$ is the charge of $j$-th particle. Fourier transform of $\rho(r)$ yields $\rho(q) = \sum_{j=1}^{N} Q_j \exp(-iqr_j)$. Sandwiching $\rho(q)$ between vectors $\langle 0| = \langle 0,0|$ and $|S\rangle = |n,k\rangle$, we get:

$$\rho_{0S}(q) = \sum_{j=1}^{N} Q_j \langle 0|e^{-iqr_j}|S\rangle \equiv \sum_{j=1}^{N} Q_j \rho_{0S}^j. \tag{S1.2.3}$$

Each term in (S1.2.3) can be written as:

$$\langle 0|e^{-iqr_j}|S\rangle = \langle 0,0|e^{-iqr_j}|n,k\rangle = A^{-1} \int \Psi_0^* \Psi_n e^{ikr_C} e^{-iqr_j} \delta\left(\sum_l \frac{m_l r'_l}{M}\right) dr'_1 dr'_2 ... dr'_N dr_C. \tag{S1.2.4}$$

Here $m_j$ is the mass of $j$-th particle, $M = \Sigma m_j$, and the delta-function represents the constraint condition $\Sigma m_l r'_l / M = 0$. Expanding the delta-function into a Fourier integral, we get:

$$\langle 0,0|e^{-iqr_j}|n,k\rangle = A^{-1} \int \Psi_0^* \Psi_n e^{ikr_C} e^{-iqr_C} e^{-iqr'_j} \left[\frac{1}{(2\pi)^D} \int e^{i\frac{m_1}{M} r'_1 p + i\frac{m_2}{M} r'_2 p + ...} dp\right] dr'_1 dr'_2 ... dr'_N dr_C. \tag{S1.2.5}$$

Here D is the dimensionality of a system, two for the case of excitons in TMDCs. Integration over $r_C$ yields

$$\langle 0,0|e^{-iqr_j}|n,k\rangle = A^{-1} \delta(k-q) \int dp \int \Psi_0^* \Psi_n e^{i\frac{m_1}{M} r'_1 p} e^{i\frac{m_2}{M} r'_2 p} ... e^{i\left(\frac{m_j}{M} p - q\right) r'_j} ... e^{i\frac{m_N}{M} r'_N p} dr'_1 dr'_2 ... dr'_N. \tag{S1.2.6}$$

The integral over all "primed" coordinates $r'_1, r'_2, ..., r'_N$ can be expressed in terms of a Fourier transform $\mathcal{F}$ of the product $\Psi_0^* \Psi_n$:

$$\langle 0,0|e^{-iqr_j}|n,k\rangle = A^{-1} \delta(k-q) \int \mathcal{F}[\Psi_0^* \Psi_n]\left(\frac{m_1}{M} p, \frac{m_2}{M} p, ..., \frac{m_j}{M} p - q, ..., \frac{m_N}{M} p\right) dp. \tag{S1.2.7}$$

Let us now plug (S1.2.7) into (S1.2.1). Presence of the delta-function $\delta(k-q)$ in (S1.2.7) reduces summation over three indices $\{n,k,q\}$ in (S1.2.1) to summation over two indices $\{n,q\}$. Then, correction to the EC ground state energy reads:

$$\Delta E_0 = -\frac{1}{2} \frac{1}{A} \sum_S |\rho_{0S}|^2 \frac{2}{\pi} \int_0^\infty \frac{\text{Im} V_s(\omega, k)}{\omega + E_{S0}} d\omega. \tag{S1.2.8}$$

Here summation is performed only over multi-index $S = \{n, k\}$. Due to presence of the delta-function in (S1.2.7) here is no need to explicitly write $\rho_{0S}(k)$, as in (S1.2.1), because momentum $k$ is already included in the multi-index $S$. Instead we simply write $\rho_{0S} = \rho_{\{0,0\},\{n,k\}} = \langle 0,0|\rho(k)|n,k\rangle$ rather than $\rho_{0S}(k)$. Since, indices $k$ and $q$ are interchangeable, (S1.2.8) can also be written in terms of the wavenumber $q$: with potential $V_s(\omega,q)$ and multi-index $S = \{n,q\}$. Such substitution yields equation (2) of the main text.

### S1.3. Selection rule for transition charge densities.

Let us now show that in certain cases transition charge densities created by electron and a hole vanish and do not contribute to the energy correction (S1.2.1, S1.2.8). Let us consider linear optical processes, when one absorbed photon creates one electron-hole pair. This photoexcited electron-hole pair can constitute a single neutral exciton or be a part of a larger excitonic complex. The expression for the transition charge density created by this photoexcied pair is written as $\rho_{0S}^{(e-h)} = \rho_{0S}^{h} - \rho_{0S}^{e}$, where $\rho_{0S}^{e}$ is the transition density for the electron and $\rho_{0S}^{h}$ is the transition density for the hole. Using equation (S1.2.7), we get (for $q = k$):

$$\rho_{0S}^{(e-h)} = (2\pi)^{-D} \int dp \left\{ \mathcal{F}[\Psi_0^*\Psi_n]\left(...,\frac{m_e}{M}p,...,\frac{m_h}{M}(p-q),...\right) - \mathcal{F}[\Psi_0^*\Psi_n]\left(...,\frac{m_e}{M}(p-q),...,\frac{m_h}{M}p,...\right) \right\}.$$
(S1.3.1)

It is easy to see that this expression vanishes if two conditions are met sumultaniously: (i) electron and hole masses are equal: $m_e = m_h = m$ and (ii) the product $\Psi_0^*\Psi_n$ is symmetric with respect to exchange between electron and hole coordinates, i.e. $\{r_e \leftrightarrow r_h\}$: $\Psi_0^*\Psi_n(..,r_e,...,r_h,..) = \Psi_0^*\Psi_n(..,r_h,...,r_e,..)$. If the two conditions are satisfied, then both terms in (S1.3.1) are identical and $\rho_{0S}^{(e-h)}$ vanishes and does not contribute to the sum in (S1.3.2).

Condition (i) can be alternatively expressed in terms of the symmetries of translational part $\exp(iq[..+m_e r_e +..+ m_h r_h +..]/M)$ of the total wavefunction $\psi_{q,n}$. If this translational part is symmetric under $\{r_e \leftrightarrow r_h\}$-exchange for any $q$, this automatically implies that electron and hole masses are equal and condition (i) is satisfied. Condition (ii) can also be expressed in terms of symmetries: it is satisfied when ground and excited state wavefunctions $\Psi_0$ and $\Psi_n$ are either both symmetric or both antisymmetric with respect to the particle exchange operator $\{r_e \leftrightarrow r_h\}$. Thus, conditions (i) and (ii) can both be expressed in terms of the symmetries of EC wavefunctions: *transition charge densities created by photoexcited electron and hole cancel each other out if both ground- and excited-state wavefunctions are either symmetric or antisymmetric with respect to exchange between electron and hole coordinates*:

$$\{r_e \leftrightarrow r_h\}\begin{pmatrix}\psi_{0,q}\\ \psi_{n,q}\end{pmatrix} = \pm\begin{pmatrix}\psi_{0,q}\\ \psi_{n,q}\end{pmatrix}.$$
(S1.3.2)

If at least one of the wavefunctions is asymmetric (i.e. neither symmetric nor antisymmetric), then electron and hole transition charge densities do not cancel each other out and their combined transition charge density does not vanish. Analogously to selection rules in atomic physics, this symmetry-based condition prohibits certain transitions in (S1.2.1, S1.2.8).

### S1.4. Charge density matrix element for unbound states.

The summation in (S1.2.8) also runs over bound as well as unbound states (i.e. states corresponding to transition energy above the binding energy $|E_{bind}|$) of an EC. It is important to understand the contribution of the unbound states to sums in (S1.2.1, S1.2.8). Suppose we have $N$ particles constituting our excitonic complex and one of them (call it particle 1) becomes unbound from the rest. That particle has momentum $p$ in the CM-frame, while the other particles have the total momentum of $-p$. Then the total wavefunction can be written as a product of wavefunction of the first particle

$$\psi_{k,p}^{1,unb.} = A^{-1/2} \exp(ikr_C + ipr_1') \qquad (S1.4.1)$$

(superscript "*unb.*" denotes unbound state) and wavefunction of the rest of the particles (remaining EC)

$$\psi_{k,p}^{N-1,unb.} = A^{-1/2} \exp(ikr_C - ipr_C^{N-1})\Psi^{N-1}(r_2', r_3',...). \qquad (S1.4.2)$$

Here $A$ is the sample volume and $r_C^{N-1} = -m_1 r_1' / M_{N-1}$ is the center-of-mass coordinate of the remaining excitonic complex with $M_{N-1} = m_2 + ... + m_N$. CM-wavefunction of the remaining excitonic complex is denoted as $\Psi^{N-1}$. Thus, the wavefuntion of an unbound EC state reads:

$$\psi_{k,p}^{unb.} = A^{-1} e^{ikr_C} e^{i(1+m_1/M_{N-1})pr_1'} \Psi^{N-1}(r_2', r_3',...) \equiv A^{-1/2} e^{ikr_C} \Psi_p^{unb.}. \qquad (S1.4.3)$$

In order to apply the approach already developed in section S1.2, we brought the wavefunction to the same form as (S1.2.2) by substituting $\Psi_p^{unb.} = A^{-1/2} e^{i(1+m_1/M_{N-1})pr_1'} \Psi^{N-1}(r_2', r_3',...)$ instead of $\Psi_n(r_1', r_2',...)$. Note that the total $N$-particle wavefunction $\psi_n$ has dimensions of $distance^{-DN/2}$ and a CM-wavefunction $\Psi_n$ in (S1.2.2) has a dimensional pre-factor $\sim a_0^{-D(N-1)/2}$, where $a_0$ is the EC Bohr radius (e.g. in case of 3D hydrogen atom, ground state wavefunction has a dimensional pre-factor $\sim a_0^{-3/2}$). At the same time, according to (S1.4.3), wavefunction $\Psi_p^{unb.}$ has a much smaller pre-factor $A^{-1/2} a_0^{-D(N-2)/2}$. Additionally, the CM-wavefunction (S1.2.2) is localized in space within characteristic length scale proportional to the effective EC Bohr radius $a_0$. This means that in the momentum space transition charge density, being a Fourier transform of the product of ground- and excited-state wavefunctions, has a characteristic width $\propto 1/a_0$. Thus, a transition charge density (S1.2.7) has maximum at small wavenumbers $q \sim 1/a_0$ and decays at large wavenumbers $q \gg 1/a_0$. In the case of transition to an unbound state, due to the factor $\exp[i(1+m_1/M_{N-1})pr_1']$ in (S1.4.3), the maximum of the corresponding transition charge density will be around wavenumber $(1+m_1/M_{N-1})p$ (since presence of the exponent inside a Fourier transform essentially shifts the resulting transformed function). At wavenumbers, significantly different from $(1+m_1/M_{N-1})p$, the transition charge density will go to zero. Thus, at large momentum $p$, transition charge density $\rho_{0,unb.}(q)$, corresponding to the unbound final state, can be treated as a broadened delta-function $\rho_{0,unb.}(q) \sim \delta(q - [1+m_1/M_{N-1}]p)$. To calculate the contribution of unbound states to (S1.2.8) we plug $\Psi_p^{unb.}$ (as defined in S1.4.3) into equation (S1.2.8). Then, the transition charge density $\rho_{0,unb.}(q)$, acting as a broadened delta-function, selects only high-momentum terms in (S1.2.8):

$$\Delta E_0(n=0 \leftrightarrow unb.) \sim \sum_q |\delta(q - [1+m_1/M_{N-1}]p)|^2 \frac{2}{\pi} \int_0^\infty \frac{\mathrm{Im} V_s(\omega,q)}{\omega + E_{S0}} d\omega \sim \tilde{V}_s(E_{S0}, p).$$

$$(S1.4.4)$$

See section S1.5 for more details regarding calculation of the frequency-integral of the screening potential $\tilde{V}_s$. At high momenta (i.e. at small spatial scales) we see more of a "bare" (i.e. unscreened) charge and hence potential approaches a coulomb-like shape $V \sim 1/p^{D-1} \to 0$.

At the same time, the characteristic magnitude of charge density matrix element $\rho_{0,unb.}$ will be smaller than $\rho_{0,n\sim1}$ (transition charge density for ground to bound state transition) by a factor of $\sim a_0^{D/2}/A^{1/2}$ since, as mentioned above, $\psi_{k,p}^{unb.}$ and $\psi_{k,n\sim1}$ have different dimensional pre-factors.

Let us denote the contributions to (S1.2.8), that correspond to transition to unbound states as $\Delta E_0 (n=0 \leftrightarrow unb.)$. Contributions to $\Delta E_0$, corresponding to transitions to bound states with $n \sim 1$ will be denoted as $\Delta E_0 (n=0 \leftrightarrow bound)$. Let us now compare the two contributions:

$$\frac{\Delta E_0 (n=0 \leftrightarrow unb.)}{\Delta E_0 (n=0 \leftrightarrow bound)} \sim \left|\frac{A^{-1/2}}{a_0^{-D/2}}\right|^2 \frac{V_s(p \gg a_0^{-1})}{V_s(p \sim a_0^{-1})} \sim \frac{a_0^D}{A} \frac{1}{V_s(p \sim a_0^{-1})} \frac{1}{p^{D-1}}. \quad (S1.4.5)$$

Let us call (S1.4.5) a "relative contribution of unbound states to the energy correction". Unbound states also have a degeneracy of $g \sim (A^{1/D})^{D-1} p^{D-1}$. Factoring in the degeneracy and assuming that for bound states screening potential is $V_s(p \sim a_0^{-1}) \sim 1/(a_0^{-1})^{D-1}$, equation (S1.4.5) yields

$$g \frac{\Delta E_0 (n=0 \leftrightarrow unb.)}{\Delta E_0 (n=0 \leftrightarrow bound)} \sim \frac{a_0}{A^{1/D}} \quad (S1.4.6).$$

Expression (S1.4.6) evaluates relative contribution of all states with momentum absolute value lying between $|p|$ and $|p| + A^{-1/D}$ (here $A^{-1/D}$ plays the role of an elementary momentum). Let us now re-write this expression per unit energy – i.e., relative contribution of unbound states with energies between $M|p|^2/2$ and $M(|p| + A^{-1/D})^2/2$:

$$\frac{g \Delta E_0 (n=0 \leftrightarrow unb.)}{\Delta E_0 (n=0 \leftrightarrow bound)} / dE_{S0} \sim \frac{1}{M|p|A^{-1/D}} \frac{a_0}{A^{1/D}} \propto \frac{1}{\sqrt{E_{S0}}}. \quad (S1.4.7)$$

Here $dE_{S0} \sim M|p|A^{-1/D}$ plays the role of elementary energy. Thus, contribution of high-energy unbound states to (S1.2.1, S1.2.8) indeed vanishes for transition energies $E_{S0}$ much higher than the binding energy $|E_{bind.}|$.

### S1.5. Integral of imaginary part of potential. Its physical significance and properties.

In this section we derive equation (3) of the main text. This equation helps to further simplify equations (S1.2.1, S1.2.8). Below we show that integral of the imaginary part of the screening potential $\tilde{V}_s = 2\pi^{-1} \int_0^\infty \text{Im} V_s(\omega, q)/(\omega + E) d\omega$, entering (S1.2.1, S1.2.8), can be expressed in terms of the real part of the screening potential which can be calculated using simple Poisson equation. We use Kramers-Kronig relations to calculate the imaginary part $\text{Im} V_s(\omega, q)$ from the real function $V_s'(\omega, q) = \text{Re} V_s(\omega, q)$:

$$\tilde{V}_s(E) = \frac{2}{\pi} \int_0^\infty d\omega \frac{1}{\pi} \int_{-\infty}^\infty d\omega' \frac{1}{\omega + E} \frac{-1}{\omega - \omega'} V_s(\omega') = -\frac{2}{\pi^2} \int_{-\infty}^\infty d\omega \frac{1}{\omega + E} \ln\left|\frac{\omega}{E}\right| V_s'(\omega).$$

(S1.5.1)

Equation 1.5.1 can be viewed as a linear transformation of $V_s(\omega)$. Let's denote this transformation as $\mathcal{T}$:

$$\tilde{V}_s(E) \equiv \mathcal{T}[V_s'(\omega)](E).$$

(S1.5.2)

Let us analyze properties of transformation $\mathcal{T}$. Since potential is a real-valued function in time domain, it is symmetric in the frequency domain: $V_s(\omega) = V_s^*(-\omega)$. Using this symmetry, we re-write (S1.5.1) as

$$\tilde{V}_s = -\frac{2}{\pi^2} \int_0^\infty d\omega \frac{1}{\omega + E} \ln\left|\frac{\omega}{E}\right| V_s'(\omega) - \frac{2}{\pi^2} \int_0^\infty d\omega \frac{1}{-\omega + E} \ln\left|\frac{\omega}{E}\right| V_s'(\omega) = -\frac{4}{\pi^2} \int_0^\infty d\omega \frac{E}{E^2 - \omega^2} \ln\left|\frac{\omega}{E}\right| V_s'(\omega).$$

(S1.5.3)

Introducing notation $\omega_{Log} = \ln \omega / \Omega$ (logarithmic frequency) and $V_s'^{Log}(\omega_{Log}) = V_s'(\Omega e^{\omega_{Log}})$ (here $\Omega$ is an arbitrary unit frequency, for example, we can choose $\Omega = 1$Hz), the integral $\tilde{V}_s$ becomes:

$$\tilde{V}_s = \frac{4}{\pi^2} \int_{-\infty}^\infty \frac{E_{Log} - \omega_{Log}}{e^{E_{Log} - \omega_{Log}} - e^{-(E_{Log} - \omega_{Log})}} V_s'^{Log}(\omega_{Log}) d\omega_{Log} \equiv \left(\frac{2}{\pi^2} \frac{E_{Log}}{\sinh E_{Log}}\right) * V_s'^{Log}(E_{Log}).$$

(S1.5.4)

Here symbol "$*$" denotes convolution. Thus, in a logarithmic scale, $\tilde{V}_s$ in equation (2) of the main text is nothing but frequency-dependent potential smoothened by a normalized bell-shaped function $2\pi^{-2} x / \sinh x$. Average value of such function is 0 and its standard deviation is ~2.2.

### *S1.6. Approximation for dynamic screening.*

Let us now further simplify (S1.2.8) and derive equation (4) of the main text. First, we re-write (S1.2.1) truncating transitions that do not contribute to it (symmetric transitions and transitions to high-energy unbound states):

$$\Delta E_0 \approx -\frac{1}{2A} \sum_q \sum_{S=S_{min}}^{S_{max}} |\rho_{0S}(q)|^2 \tilde{V}_s(E_{S0}, q).$$

(S1.6.1)

Here $S_{min}$ is lowest asymmetric state and $S_{max}$ - unbound state with corresponding transition energy of $\sim |E_{bind.}|$. As mentioned in the main text, this equation with frequency-dependent interaction potential can be re-written by replacing the transition energy $E_{S0}$ by an effective constant energy $E_{eff}$, lying between minimum and maximum bounds $E_{min}$ and $E_{max}$. Such substitution can be done based on the second mean value theorem for integrals: $\int_a^b f(\omega) g(\omega) d\omega = f(c \in [a,b]) \int_a^b g(\omega) dx$. Thus, we can pull the frequency-dependent $\tilde{V}_s$ out of the summation over $S$:

$$\Delta E_0 \approx -\frac{1}{2A}\sum_q \tilde{V}_s\left(E_{eff},q\right) \sum_{S=S_{min}}^{S_{max}} \left|\rho_{0S}(q)\right|^2 \tag{S1.6.2}$$

with $E_{min} < E_{eff} < E_{max}$.

Using definition of transition charge density (S1.2.3) and completeness of the set of state-vectors $|S\rangle$, we rewrite (S1.6.2) as:

$$\Delta E_0 = -\frac{1}{2A}\sum_q \left\{\sum_S \langle 0|\rho(q)|S\rangle\langle S|\rho(q)|0\rangle \tilde{V}_s\left(E_{eff},q\right)\right\} = \langle 0| -\frac{1}{2A}\sum_q \rho^2(q)\tilde{V}_s\left(E_{eff},q\right)|0\rangle. \tag{S1.6.3}$$

Equation (S1.6.3) is formally identical to the energy correction in the first-order perturbation theory. The expression between $\langle 0|$ and $|0\rangle$ in (S1.6.3) is nothing but the energy of a field created by charge density $\rho$ - i.e. interaction and self-action potential energies of particles constituting the EC. This energy can be also expressed as $\tilde{U}_s = \sum_{j,k}^{N} Q_j Q_k \tilde{V}_s\left(E_{eff}, r_j - r_k\right)$. Adding (S1.6.3) to the unperturbed EC energy we get the perturbed EC ground state energy:

$$E'_0 = E_0 + \Delta E_0 = \langle 0|T+U_0|0\rangle + \langle 0|\tilde{U}_s\left(E_{eff}\right)|0\rangle = \langle 0|T+\tilde{U}\left(E_{eff}\right)|0\rangle, \tag{S1.6.4}$$

where $T$ and $U_0$ are total kinetic and potential energies of unperturbed EC. Expression (S1.6.4) simply suggests that in order to evaluate binding energy of a dynamically screened EC we just need to solve a Schrodinger equation with potentials calculated with medium dielectric functions taken at energy $E_{eff}$.

# S2. Modelling of ECs in 2D systems.

## *S2.1. Analysis of interaction potential.*

In our experimental study, monolayer TMDC lies on top of the SiO$_2$ substrate (dielectric constant ~3), while environment on top of TMDC is varied. Interaction potential between two charges inside such a thin dielectric layer was derived by L. Keldysh[3]:

$$V(\rho) = \frac{\pi}{\varepsilon d}\left[H_0\left(\frac{\varepsilon_{top}+\varepsilon_{bot}}{\varepsilon d}\rho\right) - Y_0\left(\frac{\varepsilon_{top}+\varepsilon_{bot}}{\varepsilon d}\rho\right)\right]. \tag{S2.1.1}$$

Here $\rho$ is spatial separation between the particles, $\varepsilon$ and $d$ are dielectric constant and thickness of 2D material, $\varepsilon_{top}, \varepsilon_{bot}$ are dielectric constants of top and bottom (substrate) environments; $H_0, Y_0$ are Struve function and Bessel function of the second kind respectively. This equation was derived with assumption of $d$ much smaller than the separation between the charged particles, and of $\varepsilon >> \varepsilon_{top}, \varepsilon_{bot}$. These assumptions hold in some of our experimental systems, e.g. for a WS$_2$ monolayer ($\varepsilon \sim 10$) stacked between SiO$_2$ substrate ($\varepsilon_{bot} \sim 3$) and either vacuum ($\varepsilon_{top} = 1$) or ionic liquid ($\varepsilon_{top} \sim 2$ for frequencies higher than GHz). Unfortunately, for other systems these assumptions fail. Therefore, in this section we re-derive screened potentials for the geometries relevant in our experiments.

### *Heterostructure of two thin semiconductors.*

Since equation (S2.1.1) was derived with the assumption of homogeneous dielectric constant of the thin-layer material, it cannot be directly applied to stacked heterostructures of two different 2D materials (e.g. WS$_2$/MoS$_2$, WS$_2$/graphene). In case of a homogeneous monolayer material the potential $V(\rho)$ can be obtained as a solution of Poisson's equation

$$\nabla^2 V = -\delta(\vec{r}) + \chi_{2D}\delta_d(z)\nabla^2_{\vec{\rho}}V, \tag{S2.1.2}$$

where $\vec{\rho}$ is in-plane coordinate, $z$ - out-of-plane coordinate, $\vec{r} = (\vec{\rho}, z)^T$, and $\chi_{2D} = (\varepsilon - 1)d \approx \varepsilon d$ is a 2D analogue of polarizability. Note that in equation (S2.1.1) $\varepsilon$ and $d$ enter only as a product $\varepsilon d \approx \chi_{2D}$. The term $\delta(\vec{r})$ represents charge density creating the potential (we set elementary charge to 1) and broadened delta-function $\delta_d(z)$ indicates that our 2D crystal is localized in the x-y plane and has small thickness $d$. Cudazzo, Tokatly and Rubio[4] elegantly show that the solution of (S2.1.2) has exactly the same form as (S2.1.1). The 2D polarization density $\chi_{2D} = \chi_{3D}d$ characterizes dipole moment induced by electric field per unit area. In the case of a heterostructure of two monolayer materials, the 2D polarizabilities of two thin materials ($\chi^{(1)}_{2D}$ and $\chi^{(2)}_{2D}$) are additive. Thus, under the effective medium approximation, our heterostructure can be formally treated as a single (homogenious) material with effective 2D polarizability

$$\chi^{eff}_{2D} = \chi^{(1)}_{2D} + \chi^{(2)}_{2D}. \tag{S2.1.3}$$

This means that dealing with heterostructures (SiO$_2$/WS$_2$/graphene, SiO$_2$/WS$_2$/MoS$_2$, SiO$_2$/WS$_2$/WS$_2$), we can still use Keldysh equation (S2.1.1). In that equation $\varepsilon d$ must be substituted by $\chi^{eff}_{2D} \approx \varepsilon d + (\varepsilon' - 1)d'$, where $\varepsilon'$ and $d'$ are the dielectric constant and the thickness of the screening 2D material deposited on top of our WS$_2$.

### *Monolayer semiconductor between environments* $\varepsilon \approx \varepsilon_{bot} \approx \varepsilon_{top}$.

Now let us consider how a monolayer semiconductor is screened by different 3D environments. We start with the case when surrounding environments have dielectric constants similar to the dielectric constant of our 2D material. In this case, interaction potential is simply reduced to the Coulomb form:

$$V(\rho) = \frac{1}{\varepsilon \rho}. \tag{S2.1.4}$$

Interestingly, for small thickness $d$, argument of Struve and Bessel functions in (S2.1.1) is always large ($\rho/d \gg 1$) and in this limit, $H_0(x \gg 1) - Y_0(x \gg 1) \approx 2\pi^{-1} x^{-1}$. In this case, (S2.1.1) also yields a Coulomb potential $1/\varepsilon \rho$. Thus, equation (S2.1.1), derived under more strict assumptions, is still valid for the case of $\varepsilon \approx \varepsilon_{bot} \approx \varepsilon_{top}$.

### *Monolayer semiconductor between environments* $\varepsilon_{bot} \ll \varepsilon \approx \varepsilon_{top}$.

Condition of $\varepsilon_{bot} \ll \varepsilon \approx \varepsilon_{top}$ occurs, for example, when WS$_2$ is deposited on a SiO$_2$ substrate and is covered by a strongly screening 3D medium, such as ionic liquid. In this case, equation (S2.1.1) needs to be rederived. In momentum space, the potential $V(k)$ for any thickness and any set of dielectric constants reads[3]:

$$V(k) = \frac{2\pi}{\varepsilon|k|} \frac{1 + (A_1 + A_2)e^{-|k|d} + A_1 A_2 e^{-2|k|d}}{1 - A_1 A_2 e^{-2|k|d}} = \frac{4\pi}{\varepsilon|k|} \frac{\cosh\left(|k|\frac{d}{2} + \eta_2\right)\cosh\left(|k|\frac{d}{2} + \eta_1\right)}{\sinh(|k|d + \eta_1 + \eta_2)}. \tag{S2.1.5}$$

Here $A_{1,2} = (\varepsilon - \varepsilon_{bot,top})/(\varepsilon + \varepsilon_{bot,top})$ and $\eta_{1,2} = -(1/2)\ln A_{1,2}$. Substituting $\varepsilon_{bot} \ll \varepsilon \approx \varepsilon_{top}$ we get $A_2 \ll A_1 \sim 1$ and hence,

$$V(\rho) \approx \int \frac{d^2k}{(2\pi)^2} e^{ik\rho} \frac{2\pi}{\varepsilon|k|} \frac{1 + A_1 e^{-|k|d} + O(A_2)}{1 - O(A_2)} \approx \frac{1 + A_1}{\varepsilon \rho} \approx \frac{1}{\frac{\varepsilon_{top} + \varepsilon_{bot}}{2}\rho}. \tag{S2.1.6}$$

Like in the previous case, for small $d$, Coulomb-like potential (S2.1.6) can be approximated as

$$V(\rho) \approx \frac{1}{\frac{\varepsilon_{top} + \varepsilon_{bot}}{2}\rho} \approx \frac{\pi}{\varepsilon d}\left[H_0\left(\frac{\varepsilon_{top} + \varepsilon_{bot}}{\varepsilon d}\rho\right) - Y_0\left(\frac{\varepsilon_{top} + \varepsilon_{bot}}{\varepsilon d}\rho\right)\right]. \tag{S2.1.7}$$

### *Strongly screening top environment* ($|\varepsilon_{top}| \gg \varepsilon$).

When dealing with a WS$_2$/graphene heterostructure in the low-frequency regime, graphene can be treated as an ideal conductor. In this case $A_2 \approx -1$ and

$$V(\rho) = \int \frac{d^2k}{(2\pi)^2} e^{ik\rho} \frac{2\pi}{\varepsilon|k|} \frac{1 + (A_1 - 1)e^{-|k|d} - A_1 e^{-2|k|d}}{1 + A_1 e^{-2|k|d}}. \tag{S2.1.8}$$

We can analyze this expression in two limit cases:

$$V(\rho \ll d) \approx \frac{1}{\varepsilon\rho} \text{ and } V(\rho \gg d) \approx \frac{1}{\varepsilon\rho} - \frac{1}{\varepsilon\sqrt{\rho^2 + \left(\frac{1+3A_1}{1+A_1}\right)d^2}} \quad (S2.1.9)$$

This limit behavior is identical to the potential of a vertical dipole of the size $\sqrt{(1+3A_1)/(1+A_1)}d$.

For small TMDC thickness, field created by such dipole will be vanishingly small and can be written as equation (S2.1.1) putting top dielectric constant to infinity.

*Summary.*

Surprisingly, the Keldysh equation (S2.1.1) can be used for all experimentally relevant cases $\varepsilon \gg \varepsilon_{bot}, \varepsilon_{top}$; $\varepsilon_{bot} \ll \varepsilon \approx \varepsilon_{top}$; $\varepsilon_{bot} \approx \varepsilon \approx \varepsilon_{top}$; $\varepsilon_{bot} \ll \varepsilon \ll \varepsilon_{top}$, $\varepsilon_{bot} \sim \varepsilon \ll |\varepsilon_{top}|$. For heterostructures, 2D polarizability $\varepsilon d + (\varepsilon' - 1)d'$ should be substituted into (S2.1.1) instead of $\varepsilon d$ ($\varepsilon$, $d$, $\varepsilon'$, $d'$ are dielectric constants and thicknesses of materials in the heterostructure).

### S2.2. Numerical calculations.

Having justified the applicability of equation (S2.1.1) to the systems studied in this work, we performed numerical calculations of the binding energies of ECs bound by the Keldysh interaction potential.

Binding energies were calculated by variationally solving the *N*-body Schrodinger equation with the Hamiltonian defined in equation (S1.6.4) or, equivalently, equation (4) of the main text. The potential in that equation was taken in the form (S2.1.1). We used electron and hole masses of 0.45 electron mass[5,6] and WS$_2$ thickness of 0.7nm. The trial wavefunction was defined as a linear combination of correlated Gaussian basis functions. The variational parameters of the Gaussians were chosen via random trial and error, a process known as the stochastic variational method[7,8]. A finite number of parameter sets is generated, and the one which yields the lowest total energy is used to define a new correlated Gaussian which is added to the basis set. We used 40 basis functions to model neutral excitons and 400 functions to model trions and defect-bound excitons.

We calculated upper and lower bounds of EC binding energies in the following way. We evaluated the dielectric function of WS$_2$, top environment, and the bottom environments at two frequencies, $\omega_{min}$ and $\omega_{max}$, prescribed by our model. These effective dielectric constants were inserted into the Keldysh potential (S2.1.1) entering the Hamiltonian (S1.6.4). The two values of EC binding energy obtained by variational minimization of that Hamiltonian serve as upper and lower bounds for the estimated EC binding energy. For neutral exciton, the minimum frequency ($\omega_{min}$) is 130meV (transition energy between ground and first excited states) and the maximum frequency ($\omega_{max}$) is 320meV (neutral exciton binding energy). For trion, the minimum frequency is 1meV (limited by lifetime) and the maximum frequency is 30meV (characteristic trion binding energy). For defect-bound exciton, we assume 0meV effective frequency. The effective dielectric constants evaluated using this approach are shown in Table S2.1.1. The exciton binding energies calculated using these effective dielectric constants are shown in Fig.3 of the main text.

| Device structure | EC type | Neutral exciton at $\omega_{min}$ | Neutral exciton at $\omega_{max}$ | Trion at $\omega_{min}$ | Trion at $\omega_{max}$ | Defect-bound exciton at $\omega_{min}$ |
|---|---|---|---|---|---|---|
| SiO$_2$/WS$_2$ | $\varepsilon(\omega)$ | 14 | 16 | 5 | 16 | 5 |
| | $\varepsilon_{top}(\omega)+\varepsilon_{bot}(\omega)$ | 3 | 3 | 5 | 3 | 5 |
| SiO$_2$/WS$_2$/graphene | $\varepsilon(\omega)$ | 16* | 18* | 5 | 18* | 5 |
| | $\varepsilon_{top}(\omega)+\varepsilon_{bot}(\omega)$ | 3 | 3 | $\infty$ | 3 | $\infty$ |
| SiO$_2$/WS$_2$/ionic liquid | $\varepsilon(\omega)$ | 14 | 16 | 5 | 16 | 5 |
| | $\varepsilon_{top}(\omega)+\varepsilon_{bot}(\omega)$ | 4 | 4 | 7 | 4 | 21 |
| SiO$_2$/WS$_2$/MoS$_2$ | $\varepsilon(\omega)$ | 28* | 32* | 10* | 32* | 10* |
| | $\varepsilon_{top}(\omega)+\varepsilon_{bot}(\omega)$ | 3 | 3 | 5 | 3 | 5 |

**Table S2.1.1.** Effective values of dielectric functions of our 2D material (WS$_2$), and surrounding environments. The top number is the value of $\varepsilon(\omega)$ - the dielectric constant of the intermediate layer (WS$_2$ or heterostructure) – taken at minimum and maximum frequencies. The bottom number is the combined dielectric constant of 3D materials surrounding the thin layer $\varepsilon_{top}(\omega)+\varepsilon_{bot}(\omega)$ also taken at minimum and maximum frequencies. The symbol (*) indicates that the device was modelled as a heterostructure with effective dielectric constant $\varepsilon$ of the intermediate layer replaced by the sum of WS$_2$ dielectric constant and dielectric constant of another monolayer material deposited on top (MoS$_2$ or graphene at high frequencies).

# S3. Artifact analysis.

Multiple effects other than dynamic screening may, potentially, affect the EC peak positions. Some of these effects are chemical modification of the samples, screening by free carriers, and effects of strain. In this section, we analyze these effects and show they cannot account for the shifts seen in Fig.2 of the main text.

### S3.1. $WS_2$ devices in liquids. Chemical modification.

We tested possible contribution of chemical reactions between $WS_2$ and its environment to the data displayed in Fig.2 of the main text. While $WS_2$ is not known to enter chemical reactions in our conditions (vacuum, low temperatures, gate voltages ~1V)[9], such reactions may potentially affect the position of defect-related exciton peak for devices in ionic liquids. We monitored the position of the defect-related peak while first depositing the liquid onto $WS_2$, thermally cycling the device between 78K and 340K, and finally removing the liquid. We observed that the peak returns to its original position in the end of the cycle (Fig.S3.1.1). This suggests that ionic liquids do not induce significant permanent chemical changes in $WS_2$ surface.

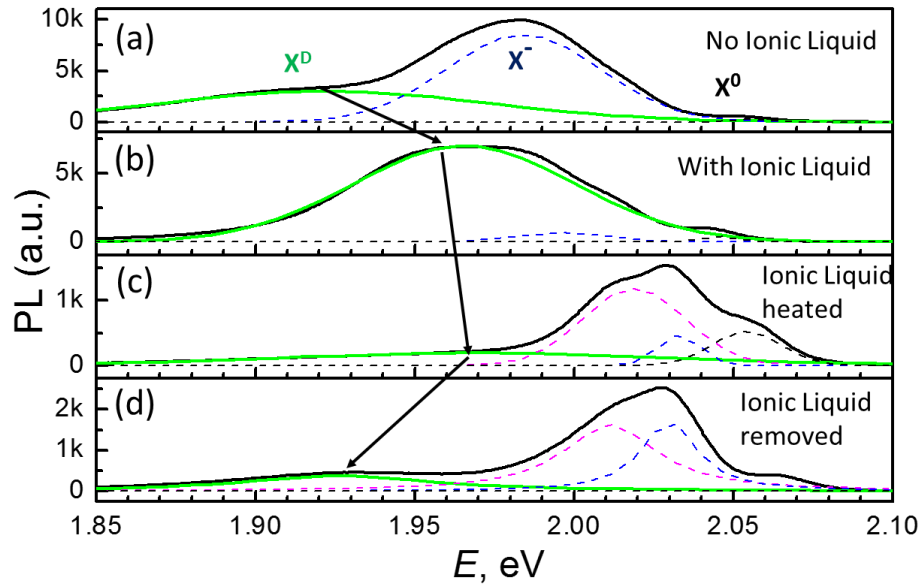

**Figure S3.1.1. Evolution of $WS_2$ photoluminescence spectra during deposition and removal of ionic liquid.** All measurements were performed at 78K temperature. **(a)** PL spectrum for a pristine $WS_2$ flake in vacuum. **(b)** PL spectrum recorded after warming the sample up to room temperature, drop-casting of ionic liquid onto it, and subsequent cooling down to 78K. **(c)** PL spectrum acquired after heating the sample to 340K, keeping it at 340K for 3 hours, and cooling it down to 78K. **(d)** PL spectrum recorded at 78K after the removal of the ionic liquid. Sample was again heated up to room temperature, ionic liquid was removed by rinsing the device with isopropanol. After that the sample was again cooled down.

Additionally, we performed experiments in which a different type of liquid – water – was deposited onto WS$_2$ samples under ambient conditions. We observed that the presence of water affects neutral and defect-bound exciton peaks in a similar way as ionic liquid (Fig. S3.1.2). Independence of the observed effects on chemical composition of the liquid confirms that peak shifts are not caused by chemical factors.

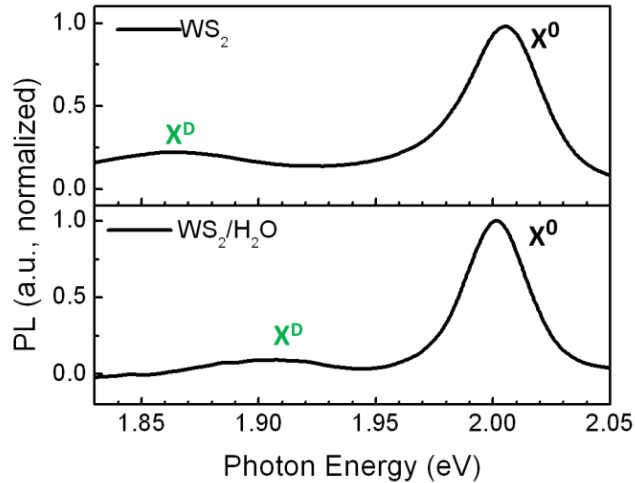

**Figure S3.1.2.** (**a**) Photoluminescence spectrum of a pristine SiO$_2$/WS$_2$ device in ambient conditions. (**b**) Photoluminescence spectrum of the SiO$_2$/WS$_2$/H$_2$O device. The presence of water causes a ~40meV blue-shift of the defect-bound exciton peak compared to the device in air. Neutral exciton does not shift, within 3meV precision.

## *S3.2. Screening by free carriers.*

Another mechanism of screening of electric interactions is screening by free carriers. Free-carrier screening causes dielectric functions to become wavenumber-dependent. In our measurements, we deal with two types of screening with different characteristic spatial scales:

(a) <u>In WS$_2$/graphene heterostructures, the fields of ECs in WS$_2$ are screened by relativistic carriers in graphene.</u> Due to absence of the bandgap and linear dispersion charge carriers, the only characteristic spatial scale in graphene is its lattice constant $a$~0.3nm. Then, the Thomas-Fermi screening length in graphene is also ~0.3nm[10]. At smaller spatial scales (larger momenta) screening is akin to weak screening by individual atoms and effective dielectric constant is $\varepsilon(q<1/a)\sim 1$, while screening becomes stronger at momenta smaller than $1/a$. Hence, for ECs in WS$_2$ with the effective Bohr radii 1~2nm graphene can be treated as a strongly screening metal, at least in the regime of low frequencies[10,11].

(b) <u>The electric fields of ECs in WS$_2$ can also be screened by free carriers in WS$_2$ itself</u>. Presence of free carriers in the material is typically caused by doping (intrinsic or induced). To investigate the effects of free-carrier screening we fabricated back-gated WS$_2$ samples on Si/SiO$_2$ substrate (with 300nm SiO$_2$ thickness) with controllable doping level. Experimentally, our data (Fig.S3.2.1) as well as the data obtained by other groups[12,13] suggest that ECs are relatively weakly affected by the presence of free carriers. Neutral exciton peak shifts from 2.055eV in a depleted sample to 2.075eV in a doped sample (Fig.S3.2.1). The position of the defect-bound exciton peak shifts by <5meV due to gating (Fig.S2.3.1). In depleted samples the binding energy of trions is the smallest, 24meV. The trion peak experiences red-shifts in positively gated devices. This reflects the increase of the trion binding energy in presence of doping. Peak shifts observed in our experiments, described in the main text, exceed shifts that can be induced by doping: Neutral exciton red-shift to ~2.045eV in bilayer $WS_2$ samples (Fig.3.3.1) and in $WS_2/MoS_2$ heterostructures. Defect-related peaks blue-shift by ~40meV in presence of graphene or liquid environments. Trion binding energy becomes as low as 19meV in presence of graphene. Thus, screening by electron gas in 2D semiconductor cannot account for peak shifts observed in our experiments.

Relatively weak effects of free-carrier screening are also expected theoretically. In the case of screening by gas of massive 2D electrons, effective dielectric function depends on the wavenumber or, in other words, on the spatial scale at which interactions occur: dielectric function becomes large only at interparticle distances greater than characteristic screening length. The screening length lies between the exciton Bohr radius $a_0 > 1nm$ and inverse Fermi momentum $k_F^{-1}$ [14]. For our typical $10^{12}cm^{-2}$ doping level, inverse Fermi momentum is $k_F^{-1} \sim 5nm > a_0$. In this regime the material dielectric function is[14]:

$$\varepsilon(q) = \varepsilon_0 \left(1 + \frac{g/a_0}{q}\left\{1 - \sqrt{1 - \left[\frac{2k_F}{q}\right]^2}\right\}\right) \approx \varepsilon_0 \left(1 + \frac{g/a_0}{q}\frac{2k_F^2}{q^2}\right).$$

Here $\varepsilon_0$ is the dielectric constant at large momenta (small spatial scales), $g = 4$ is a spin- and valley- degeneracy factor and $q$ - wavenumber. Thus, in the case of a *moderate doping* ($10^{12}cm^{-2}$) at small spatial scales corresponding to exciton size (1~2nm), the 2D material dielectric function is close to $\varepsilon_0$ (dielectric function in absence of free carriers).

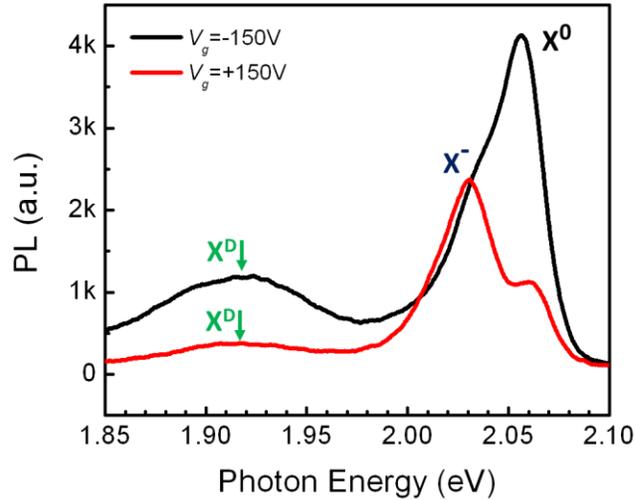

**Figure S3.2.1. Photoluminescence spectra of a gated $WS_2$ device** acquired at two different gate voltages, $V_g$=-150V and $V_g$=+150V.

## S3.3. Effects of strain.

Mechanical strain may be induced in $WS_2$ during transfer of another material onto it. In principle, strain may affect EC peak positions[15]. To roughly estimate the magnitude of that strain, we compared PL spectra of $WS_2$ in a mechanically transferred $WS_2/MoS_2$ heterostructure and of natural $WS_2$ bilayer which did not undergo any transfer (Fig.S.3.3.1). While screening in these structures is very similar, the transfer-related strain may only be present in $WS_2/MoS_2$ device. The comparison of $X^0$ peak positions between the spectra of $WS_2/MoS_2$ heterostructure and natural bilayer $WS_2$ shows only ~2meV difference and suggests that the stain imparted by the transfer is smaller than ~0.03%[15]. This strain and corresponding peak shift is too small to account for the trends seen in Fig.2 of the main text.

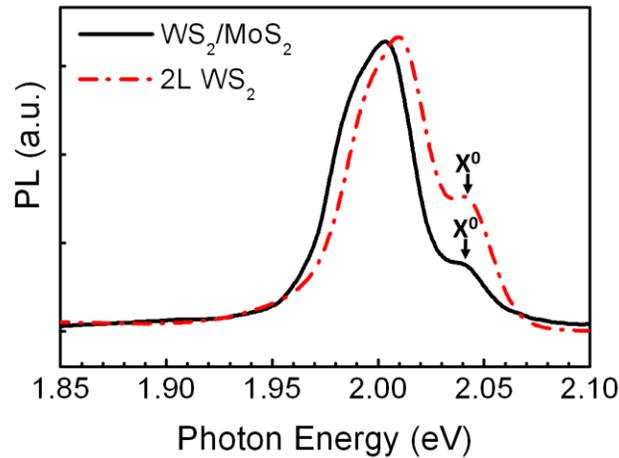

**Figure 3.3.1. Photoluminescence spectra of artificial $WS_2/MoS_2$ heterostructure and natural bilayer $WS_2$.**